Growth and anisotropy evaluation of NbBi$Ch_3$ ($Ch$ = S, Se) misfit-layered superconducting single crystals

Masanori Nagao[a,d*], Akira Miura[b], Yoichi Horibe[c], Yuki Maruyama[a], Satoshi Watauchi[a], Yoshihiko Takano[d] and Isao Tanaka[a]

[a]*University of Yamanashi, 7-32 Miyamae, Kofu, Yamanashi 400-8511, Japan*

[b]*Hokkaido University, Kita-13 Nishi-8, Kita-ku, Sapporo, Hokkaido 060-8628, Japan*

[c]*Kyushu Institute of Technology, 1-1 Sensui-cho, Tobata, Kitakyushu, Fukuoka 804-8550, Japan*

[d]*National Institute for Materials Science, 1-2-1 Sengen, Tsukuba, Ibaraki 305-0047, Japan*

[*]Corresponding Author

Masanori Nagao

Postal address: University of Yamanashi, Center for Crystal Science and Technology Miyamae 7-32, Kofu 400-8511, Japan

Telephone number: (+81)55-220-8610

Fax number: (+81)55-254-3035

E-mail address: mnagao@yamanashi.ac.jp

**Abstract**

NbBi$Ch_3$ ($Ch$ = S, Se) misfit-layered superconducting single crystals were successfully grown using a CsCl/KCl flux for the first time. The obtained crystals had a well-developed habit parallel to the $c$-plane with a typical width of 1–2 mm and thickness of 10–40 μm. The superconducting transition temperatures with zero resistivity of NbBiS$_3$ single crystals obtained from the nominal composition of Nb$_{0.9}$Bi$_{1.2}$S$_3$ was 0.31 K, and that value of the NbBiSe$_3$ single crystals grown from the stoichiometry composition (NbBiSe$_3$) was 2.3 K. Sharp decreases in electric resistivity and magnetic susceptibility at approximately 3 K suggested a possible superconducting transition temperature of NbBiSe$_3$. The normal-state anisotropy values of grown NbBiS$_3$ and NbBiSe$_3$ single crystals were 2.2–2.4 and 1.5–1.6, respectively.

**Main text**

**1. Introduction**

Misfit layered compounds (MLC) exhibit unique physical properties due to incommensurate layered structures [1]. The general chemical formula of these materials is represented as $(MX)_{1+\delta}(TX_2)$, where $MX$ and $TX_2$ ($M$ = Sn, Pb, Bi, Sb, rare earth elements; $T$ = transition metal; $X$ = S, Se, Te) are monochalcogenide and transition metal dichalcogenide layers, respectively [2-4]. In particular, the physical properties of the incommensurate direction have attracted much attention. No simple periodic structure will be expected on the incommensurate direction; furthermore, band calculation may not be effective for the prediction of physical properties. Investigation of these compounds along the incommensurate direction will provide a new possibility in physics. Some MLCs become superconductors [4-7], and their transport properties along the incommensurate direction are more interesting. Moreover, we hope these experimental results can apply to quasicrystals [8], which have complex periodic structures. In focusing on the application, MLC are promising candidates for the improvement in the performance of thermoelectric materials [9-11].

A major challenge is the control of misfit layered compound structures via simple synthesis techniques. The single crystals and thin films of these compounds have often been grown using the chemical vapor transport (CVT) [12,13] and the physical vapor deposition (PVD) method [14]. This technique enables the control of "layer by layer" and new metastable structures have been obtained. However, severe control of growth conditions is required and the growth rate is low. As a result, measuring physical properties perpendicular to the layers is difficult. In contrast, the flux method is a popular technique for growing layered chalcogenides [15,16] and the growth rate is higher than that of CVT and PVD method. The flux method may be a potential method for the growth of bulk single crystals to measure the physical properties perpendicular to the layers. In this letter, we have successfully grown single crystals of NbBi$Ch_3$ ($Ch$ = S, Se) MLC using a CsCl/KCl flux. These NbBi$Ch_3$ ($Ch$ = S, Se) are superconductors with transition temperatures of 0.3 and 2.3 K, respectively [6]. The superconducting property and normal state transport anisotropy of NbBi$Ch_3$ ($Ch$ = S, Se) were investigated using the obtained single crystals.

## 2. Experimental

The NbBi$Ch_3$ ($Ch$ = S, Se) single crystals were grown using a CsCl/KCl flux [17]. The raw materials of Nb, Bi, and $Ch$ were weighed with a nominal composition of

NbBi$Ch_3$ and Nb$_{0.9}$Bi$_{1.2}$$Ch_3$. The molar ratio of the flux was CsCl:KCl = 5:3. The mixture of the raw materials (0.8 g) and the flux (5.0 g) was ground using a mortar and then sealed into an evacuated quartz tube (~10 Pa). The quartz tube was heated at 700 °C for 10 h and subsequently cooled to 600 °C at a rate of 0.5 °C/h. The sample was then spontaneously cooled to less than 30 °C in the furnace. The resulting quartz tube was opened in air, and the obtained products were washed and filtered using distilled water to remove the CsCl/KCl flux.

The compositional ratio of the grown single crystals was evaluated using energy dispersive X-ray spectrometry (EDS) (Bruker, Quantax 70), and the microstructure was observed using a scanning electron microscope (SEM) (Hitachi High-Technologies, TM3030). The obtained compositional values were normalized using $Ch$ = 3.00. The impurity elements from CsCl/KCl flux were determined by electron probe microanalysis with wavelength dispersive X-ray spectrometry (WDS) (EPMA, JXA-8200, JEOL). The identification and orientation of the grown crystals were performed through X-ray diffraction (XRD) using Rigaku MultiFlex with CuK$\alpha$ radiation.

The temperature dependence of magnetic susceptibility ($M$-$T$ curve) under zero-field cooling (ZFC) and field cooling (FC) with 10 Oe was measured using a

superconducting quantum interface device (SQUID) with an applied magnetic field parallel to the *c*-axis. The critical temperature ($T_c$) of the obtained single crystals was estimated from the *M-T* curve.

The resistivity-temperature (*ρ-T*) characteristics of the single crystals were measured using the standard four-probe method in the constant current (*J*) mode with a physical property measurement system (PPMS; Quantum Design DynaCool). Measurement of the *ρ-T* characteristics below 1.8 K was performed using an adiabatic demagnetization refrigerator (ADR) option on the PPMS. The magnetic field applied for operating the ADR was 3 T at 1.9 K; subsequently, it was removed. Consequently, the temperature of the sample decreased to approximately 0.15 K. The measurement of the *ρ-T* characteristics was commenced at the lowest temperature (approximately 0.15 K), which was spontaneously increased to 15 K. The electrical terminals were made of silver paste. The transition temperature corresponding to the onset of superconductivity ($T^{onset}_c$) was defined as the point at which deviation from linear behavior is observed in the normal conducting state in the *ρ-T* characteristics. The zero resistivity ($T_c^{zero}$) was determined considering the criterion of a resistivity of 1.0 μΩcm in the *ρ-T* characteristics. The obtained single crystals were fixed on the MgO single-crystalline substrate for the *c*-axis transport measurement, and they were subsequently fabricated to

s-shaped junctions using a three-dimensional (3D) focused Ga-ion beam (FIB) etching method [18,19]. The area of the junctions was approximately 3.8–5.8 μm × 2.9–3.7 μm in the *c*-plane. The thickness of the junctions was approximately 0.9–1.5 μm along the *c*-axis. Normal-state anisotropy ($\gamma_n$) was estimated from the *ρ-T* characteristics of the as-grown (*J*//*c*-plane) and fabricated (*J*//*c*-axis) samples.

**3. Results and discussion**

We successfully grew four types of single crystals from different starting materials and ratios. The obtained single crystals were thin plates with width and thickness in the ranges 1–2 mm and 10–40 μm, respectively (See Figure S1 in Supplementary Materials). However, the crystals grown from $NbBiS_3$ starting materials were smaller than other crystals. The estimated compositions of the obtained single crystals are shown in Table I. All the crystals were close to the stoichiometric ratio Nb:Bi:*Ch*=1:1:3 although the crystals grown from $Nb_{0.9}Bi_{1.2}Se_3$ was slightly Nb-poor and Bi-rich. Cs, K, and Cl from the flux were not detected in the obtained single crystals by qualitative analysis using WDS with a minimum sensitivity limit of approximately 0.1 wt%. Further examination was performed on three types of crystals grown from $Nb_{0.9}Bi_{1.2}S_3$, $NbBiSe_3$, and

Nb$_{0.9}$Bi$_{1.2}$Se$_3$ because the crystals grown from NbBiS$_3$ were small and were little amount to measure their physical properties.

Figure 1 shows the XRD patterns of a well-developed plane in the obtained single crystals grown from the nominal compositions of (a) Nb$_{0.9}$Bi$_{1.2}$S$_3$ and (b) NbBiSe$_3$ and (c) Nb$_{0.9}$Bi$_{1.2}$Se$_3$. The presence of only the 00$l$ diffraction peaks indicates that the $ab$-plane was well-developed, which corresponded to NbBiS$_3$ and NbBiSe$_3$ structures [2,20]. The $c$-axis lattice constants of NbBiS$_3$ and NbBiSe$_3$ were approximately 22.97 and 24.18 Å, respectively. These $c$-axis lattice constants were similar to those of previous reports [6,13]. In NbBiSe$_3$, the XRD patterns of the obtained single crystals in the nominal compositions between NbBiSe$_3$ and Nb$_{0.9}$Bi$_{1.2}$Se$_3$ were comparable.

Figure 2 shows the $\rho_{c\text{-plane}}$-$T$ characteristics along the $c$-plane for single crystals grown from a starting powder with a nominal composition of (a) Nb$_{0.9}$Bi$_{1.2}$S$_3$ and (b) NbBiSe$_3$ and Nb$_{0.9}$Bi$_{1.2}$Se$_3$. These samples exhibited metallic behavior above the superconducting transition temperature. The $\rho_{c\text{-plane}}$-$T$ behaviors of the Nb$_{0.9}$Bi$_{1.2}$S$_3$ and NbBiSe$_3$ samples were similar, but that of Nb$_{0.9}$Bi$_{1.2}$Se$_3$ exhibited a weak hump at between 50 and 150 K. The origin of that hump structure is unknown. One possibility is a structural phase transition. XRD measurement at low temperature may reveal to this phenomena. The zero-resistivity temperatures ($T_c^{\text{zero}}$) of (a) Nb$_{0.9}$Bi$_{1.2}$S$_3$ and (b)

NbBiSe$_3$ samples were estimated to be 0.31 and 2.3 K respectively. These superconducting transition temperatures corresponded to the NbBiS$_3$ and NbBiSe$_3$ phases from a previous report [6]. By contrast, $T_c^{zero}$ of Nb$_{0.9}$Bi$_{1.2}$Se$_3$ sample was approximately 3.2 K. This value was not consistent with the previous report, which was 2.36 K [6]. This transition phase may be a new superconducting phase.

We examined the anisotropic properties of the grown crystals using s-shaped junctions [21]. The s-shaped junctions on the obtained single crystals along the $c$-axis were fabricated using FIB etching (See Figure S2 in Supplementary Materials). The direction of current flow was along the $c$-axis in the fabricated region. Figure 3 shows the $\rho_{c\text{-axis}}$-$T$ characteristics along the $c$-axis for the obtained single crystals from a nominal composition of (a) Nb$_{0.9}$Bi$_{1.2}$S$_3$ and (b) NbBiSe$_3$ and Nb$_{0.9}$Bi$_{1.2}$Se$_3$. These $\rho_{c\text{-axis}}$-$T$ curves along the $c$-axis above the superconducting transition temperatures were similar to those of the $c$-plane. The $T_c^{zero}$ along the $c$-axis of (a) Nb$_{0.9}$Bi$_{1.2}$S$_3$, (b) NbBiSe$_3$ and Nb$_{0.9}$Bi$_{1.2}$Se$_3$ were determined to be 0.21, 2.2, and 2.1 K, respectively. The superconducting transition temperatures slightly decreased compared with those of the $c$-plane. That reason can be attributed to the damage caused by FIB etching. [21] The normal-state anisotropies ($\gamma_n$) were evaluated from the ratio of the resistivity using the following equation:

$$\gamma_n = (\rho_{c\text{-axis}} / \rho_{c\text{-plane}})^{1/2} \qquad (1)$$

The normal-state anisotropies ($\gamma_n$) were observed to be 2.2–2.4, 1.5–1.6, and 3.0–4.9 for $Nb_{0.9}Bi_{1.2}S_3$, $NbBiSe_3$, and $Nb_{0.9}Bi_{1.2}Se_3$ crystals, respectively.

The $\rho_{c\text{-axis}}$-$T$ characteristics along the $c$-axis of $Nb_{0.9}Bi_{1.2}Se_3$ crystal exhibited a decrease in resistivity at approximately 3 K by approximately 20%. This suggested the co-existence of an intergrowth phase with a transition temperature higher than that reported for superconducting $NbBiSe_3$ (~2 K). The $\gamma_n$ of $Nb_{0.9}Bi_{1.2}Se_3$ sample was higher than that of the $NbBiSe_3$ single crystals. This result strongly suggested that the intergrowth phase existed in the $Nb_{0.9}Bi_{1.2}Se_3$ sample. However, the intergrowth layers (phases) in these single crystals have not yet been observed using transmission electron microscopy (TEM), of which further investigation is required.

We measured $M$-$T$ curve for single crystals grown from a starting powder with a nominal composition of $NbBiSe_3$ and $Nb_{0.9}Bi_{1.2}Se_3$ under ZFC and FC (See Figure S3 in Supplementary Materials). The $T_c$ of $NbBiSe_3$ and $Nb_{0.9}Bi_{1.2}Se_3$ estimated from the separation temperature of FC and ZFC curves were 2.6 and 3.4 K, respectively. While the transition of $NbBiSe_3$ was sharp, that of $Nb_{0.9}Bi_{1.2}Se_3$ was broad. Considering misfit compounds allow different stacking layers, we considered that the phase of $Nb_{0.9}Bi_{1.2}Se_3$ partially contained the superconducting phase up to 3.4 K. On the other

hand, the *M-T* curve of the single crystal grown from the nominal compositions of Nb$_{0.9}$Bi$_{1.2}$S$_3$ exhibited no superconductivity and no specific anomaly between 2 and 10 K.

Additionally, the $\rho_{c\text{-plane}}$–*T* characteristics under magnetic fields of the obtained single crystals grown from the starting materials with a nominal composition of NbBiSe$_3$ and Nb$_{0.9}$Bi$_{1.2}$Se$_3$ were measured, and were then estimated to the upper critical fields ($H_{c2}$). The field dependences of $T_c^{\text{onset}}$ and $T_c^{\text{zero}}$ under the magnetic fields (*H*) parallel to the *c*-plane (*H* // *c*-plane) and *c*-axis (*H* // *c*-axis) for (a) NbBiSe$_3$ and (b) Nb$_{0.9}$Bi$_{1.2}$Se$_3$ samples are plotted in Figure 4. In (a) for the NbBiSe$_3$ sample, the linear extrapolations of $T_c^{\text{onset}}$ for the cases of *H* // *c*-plane and *H* // *c*-axis approached values of 2.67 and 0.88 T, respectively. The upper critical fields $H_{c2}^{//c\text{-plane}}$ and $H_{c2}^{//c\text{-axis}}$ were predicted to be less than 2.67 and 0.88 T, respectively. Using linear fitting to the $T_c^{\text{zero}}$ data, the irreversibility fields $H_{\text{irr}}^{//c\text{-plane}}$ and $H_{\text{irr}}^{//c\text{-axis}}$ were observed to be less than 0.94 and 0.46 T, respectively. Moreover, in (b) for the Nb$_{0.9}$Bi$_{1.2}$Se$_3$ sample, $H_{c2}^{//c\text{-plane}}$ and $H_{c2}^{//c\text{-axis}}$ were estimated to be less than 1.64 and 0.69 T, respectively. $H_{\text{irr}}^{//c\text{-plane}}$ and $H_{\text{irr}}^{//c\text{-axis}}$ were observed to be less than 0.85 and 0.53 T, respectively. The upper critical fields ($H_{c2}$) of the obtained single crystals grown from the nominal composition of Nb$_{0.9}$Bi$_{1.2}$Se$_3$ were lower than that of NbBiSe$_3$, even though $T_c$ of the Nb$_{0.9}$Bi$_{1.2}$Se$_3$

sample was higher than that of $NbBiSe_3$. The $NbBiSe_3$ and $Nb_{0.9}Bi_{1.2}Se_3$ samples exhibited different behaviors in their superconducting properties. Additionally, in a conventional (BCS-like) superconductor for the weak-coupling limit, the Pauli limit is $H_p = 1.84T_c$ [22], which was calculated to be 4.78 and 6.26 T for the obtained single crystals. Thus, the $H_{c2}$s in the magnetic fields parallel to the $c$-plane ($H_{c2}^{//c\text{-plane}}$ = 1.64-2.67 T) were lower than the Pauli limit ($H_p$ = 4.78–6.26 T), indicating the possibility of a conventional superconductor. In contrast, the irreversibility fields under the magnetic field applied parallel to the $c$-axis ($H_{irr}^{//c\text{-axis}}$) exhibited the opposite relation compared with the $H_{c2}$ between $NbBiSe_3$ and $Nb_{0.9}Bi_{1.2}Se_3$ samples. This result indicated the possibility that the obtained single crystals grown from the nominal composition of $Nb_{0.9}Bi_{1.2}Se_3$ had an intergrowth phase, which behaved role of a pinning site. This result of the upper critical fields indicated that the Nb-Bi-Se-based superconductors possessed anisotropy. The superconducting anisotropies ($\gamma_s$) were evaluated from the ratio of the upper critical field ($H_{c2}$) using the following equation:

$$\gamma_s = H_{c2}^{//c\text{-plane}} / H_{c2}^{//c\text{-axis}} = \xi_{c\text{-plane}} / \xi_{c\text{-axis}} \quad (\xi : \text{coherence length}) \qquad (2)$$

The superconducting anisotropies ($\gamma_s$) were 3.03 and 2.38 for the $NbBiSe_3$ and $Nb_{0.9}Bi_{1.2}Se_3$ samples, respectively.

## 4. Conclusion

NbBi$Ch_3$ (*Ch*: S, Se) misfit-layered superconducting single crystals with sizes and thicknesses in the ranges 1–2 mm and 10–40 μm, respectively, were successfully grown using a CsCl/KCl flux for the first time. The zero-resistivity temperature ($T_c^{zero}$) of the NbBiS$_3$ single crystals obtained from the nominal composition of Nb$_{0.9}$Bi$_{1.2}$S$_3$ was 0.31 K, and that temperature of the NbBiSe$_3$ single crystals grown from the stoichiometry composition (NbBiSe$_3$) was 2.3 K. Those values were consistent with the previous report. The superconducting anisotropy ($\gamma_s$) value of NbBiSe$_3$ single crystals was 3.03, and the normal-state anisotropy ($\gamma_n$) values of NbBiS$_3$ and NbBiSe$_3$ single crystals were 2.2–2.4 and 1.5–1.6, respectively. Moreover, the superconducting transition of NbBiSe$_3$ crystals grown from the nominal composition of Nb$_{0.9}$Bi$_{1.2}$Se$_3$ exhibited 3 and 2.1 K with two-step resistivity declines. The anisotropies of superconductivity ($\gamma_s$) and normal-state ($\gamma_n$) were 2.38 and 3.0–4.9, respectively. This suggests the possible existence of a superconducting misfit phase with the transition temperature of approximately 3 K, and further examination is necessary for this intergrowth phase.

**Acknowledgments**

This work was supported by Grants-in-Aid for Scientific Research (C) (JSPS KAKENHI Grant Number JP19K05248).

The authors would like to thank Mr. N. Kataoka, Prof. T. Yokoya, Prof. K. Kobayashi (Okayama University) and Prof. M. Ishimaru (Kyushu Institute of Technology) for experimental supports and useful discussions.

TEM studies were partially performed through "Advanced Characterization Nanotechnology Platform, Nanotechnology Platform Program of the Ministry of Education, Culture, Sports, Science and Technology (MEXT), Japan" at the Research Center for Ultra-High Voltage Electron Microscopy (Nanotechnology Open Facilities) in Osaka University.

We would like to thank Editage (www.editage.com) for English language editing.

Table I Chemical ratios of the obtained single crystals grown from each nominal composition. Normalized by $Ch = 3.00$.

| Nominal composition | | $Ch$:S | | $Ch$:Se | |
|---|---|---|---|---|---|
| | | NbBiS$_3$ | Nb$_{0.9}$Bi$_{1.2}$S$_3$ | NbBiSe$_3$ | Nb$_{0.9}$Bi$_{1.2}$Se$_3$ |
| Chemical ratio | Nb | 0.99(2) | 1.00(3) | 0.98(3) | 0.93(2) |
| | Bi | 1.02(3) | 1.00(1) | 0.98(2) | 1.14(5) |
| | $Ch$ | 3.00 | 3.00 | 3.00 | 3.00 |

**Figure captions**

Figure 1. XRD pattern of a well-developed plane for obtained single crystals grown from the nominal compositions of (a) $Nb_{0.9}Bi_{1.2}S_3$, (b) $NbBiSe_3$ and (c) $Nb_{0.9}Bi_{1.2}Se_3$

Figure 2. Temperature dependence of the resistivity along the $c$-plane ($\rho_{c\text{-plane}}$–$T$ characteristics) for single crystals grown from the nominal compositions of (a) $Nb_{0.9}Bi_{1.2}S_3$, (b) $NbBiSe_3$ and $Nb_{0.9}Bi_{1.2}Se_3$. The inset shows the $\rho_{c\text{-plane}}$–$T$ characteristics at around superconducting transition temperature.

Figure 3. Temperature dependence of the resistivity along the $c$-axis ($\rho_{c\text{-axis}}$–$T$ characteristics) for single crystals grown from the nominal compositions of (a) $Nb_{0.9}Bi_{1.2}S_3$, (b) $NbBiSe_3$ and $Nb_{0.9}Bi_{1.2}Se_3$. The inset shows the $\rho_{c\text{-axis}}$–$T$ characteristics at around superconducting transition temperature.

Figure 4. Field dependences of $T_c^{onset}$ and $T_c^{zero}$ under magnetic fields ($H$) parallel to the $ab$-plane ($H \mathbin{/\mkern-6mu/} ab$-plane) and $c$-axis ($H \mathbin{/\mkern-6mu/} c$-axis) for a nominal composition of (a) $NbBiSe_3$ and (b) $Nb_{0.9}Bi_{1.2}Se_3$. The lines are linear fits to the data. The inset is an enlargement of the lower-field region.

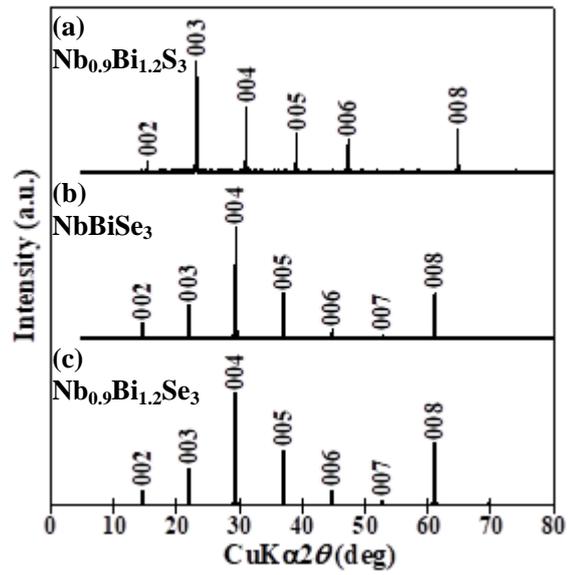

**Figure 1**

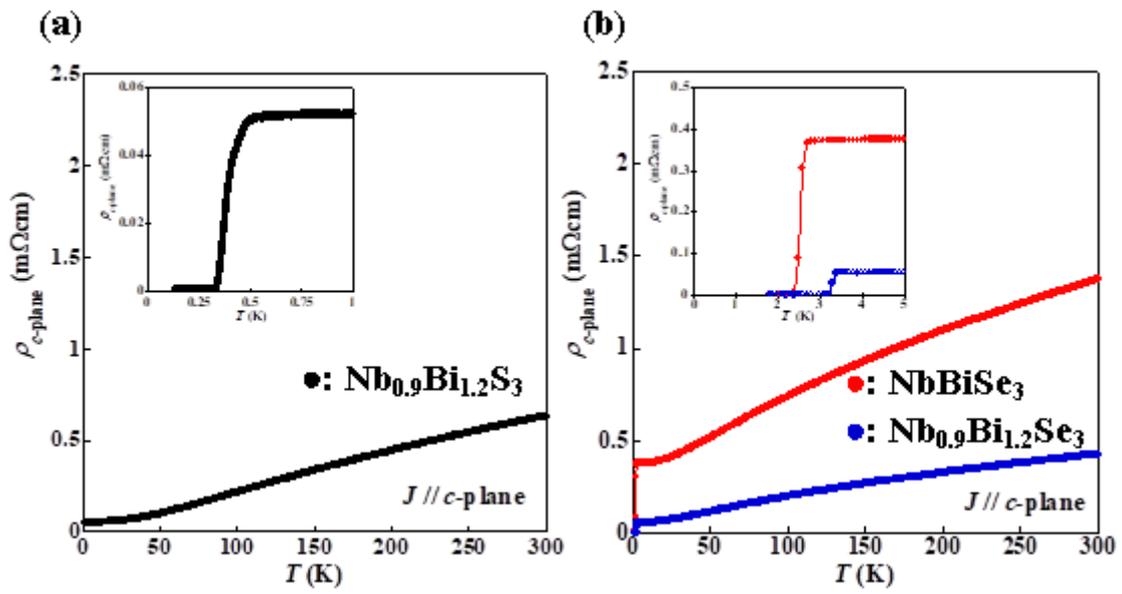

**Figure 2**

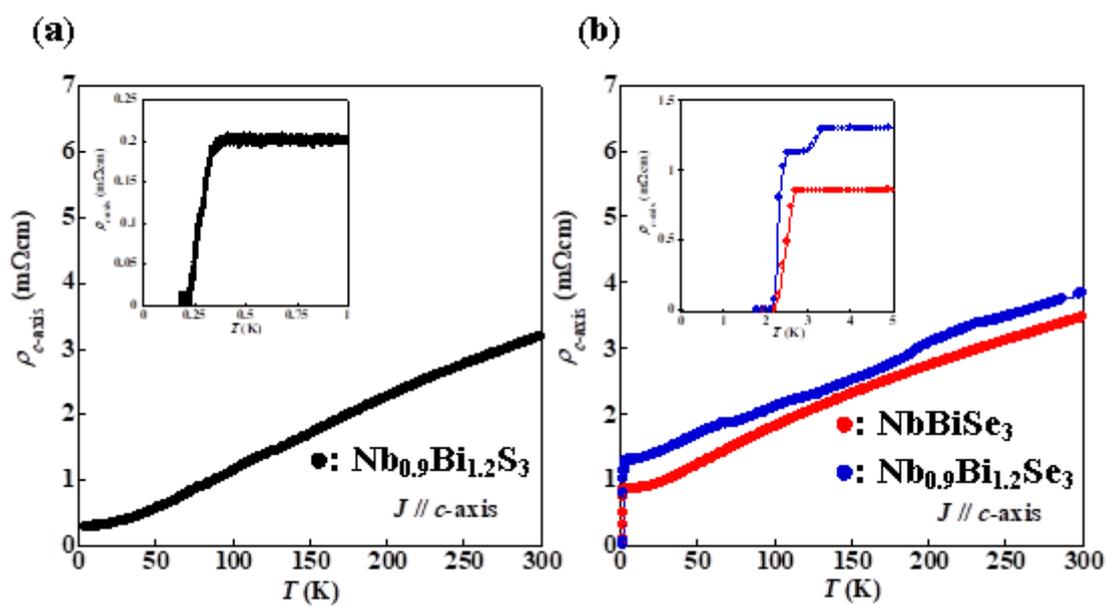

**Figure 3**

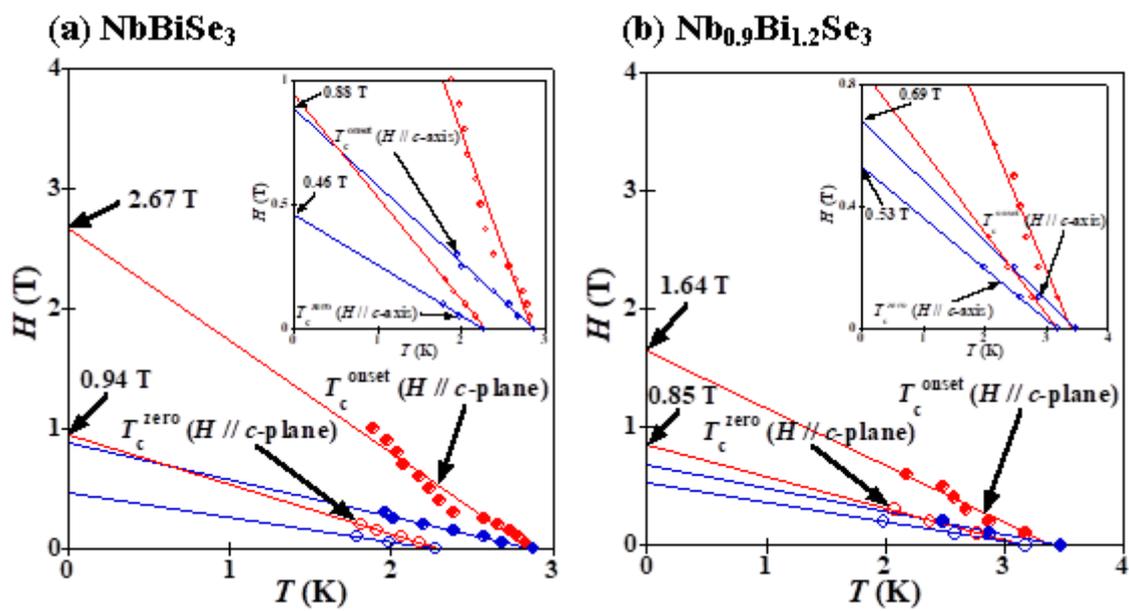

**Figure 4**

Supplementary Materials

for

**Growth and anisotropy evaluation of NbBi$Ch_3$ ($Ch$ = S, Se)**

**misfit-layered superconducting single crystals**

Masanori Nagao[a,d*], Akira Miura[b], Yoichi Horibe[c], Yuki Maruyama[a], Satoshi Watauchi[a],

Yoshihiko Takano[d] and Isao Tanaka[a]

[a]*University of Yamanashi, 7-32 Miyamae, Kofu, Yamanashi 400-8511, Japan*

[b]*Hokkaido University, Kita-13 Nishi-8, Kita-ku, Sapporo, Hokkaido 060-8628, Japan*

[c]*Kyushu Institute of Technology, 1-1 Sensui-cho, Tobata, Kitakyushu, Fukuoka 804-8550, Japan*

[d]*National Institute for Materials Science, 1-2-1 Sengen, Tsukuba, Ibaraki 305-0047, Japan*

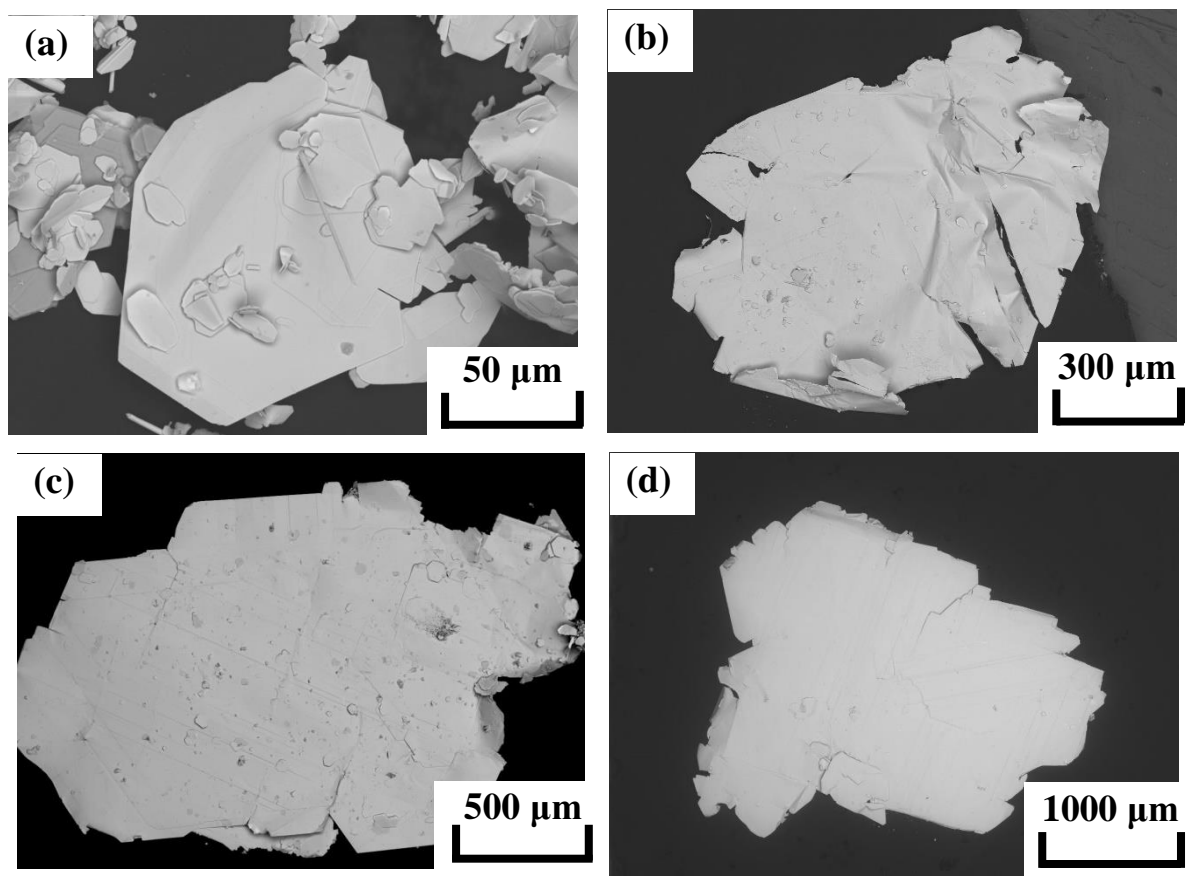

Figure S1. Typical scanning electron microscope (SEM) image of obtained single crystals grown from the nominal compositions of (a) $NbBiS_3$, (b) $Nb_{0.9}Bi_{1.2}S_3$, (c) $NbBiSe_3$ and (d) $Nb_{0.9}Bi_{1.2}Se_3$.

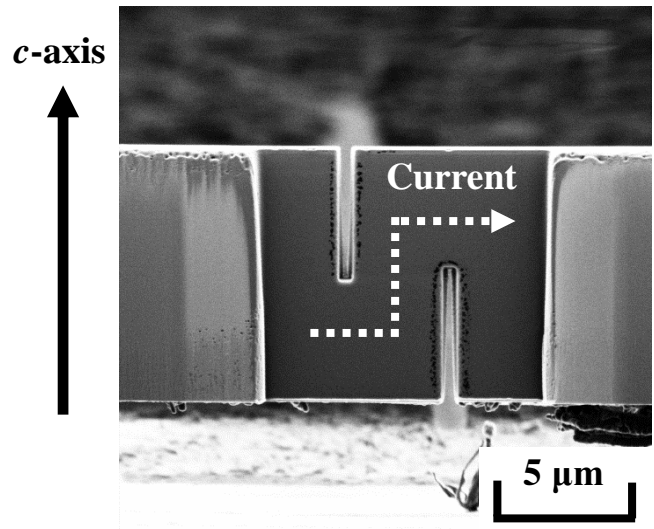

Figure S2. Scanning ion microscopy (SIM) image of a typical s-shaped junction.

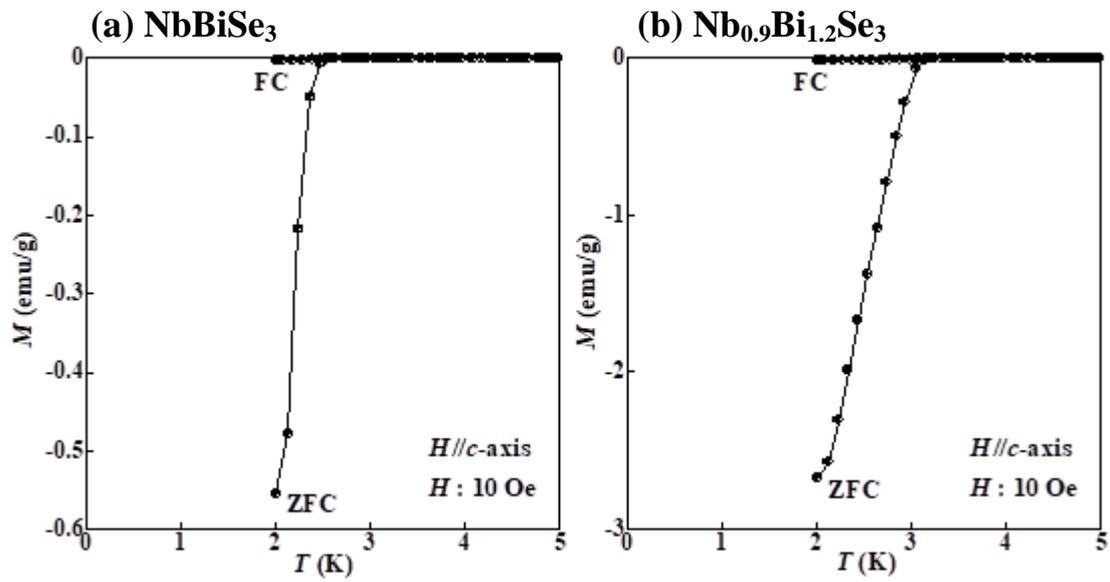

Figure S3. Temperature dependence of magnetic susceptibility (*M-T*) for obtained single crystals grown from the nominal compositions of (a) $NbBiSe_3$ and (b) $Nb_{0.9}Bi_{1.2}Se_3$ under ZFC and FC.